\documentclass[a4paper]{svmult}
\usepackage{times}
\usepackage[dvips]{graphicx,epsfig}
\usepackage{amsmath,amsfonts,amssymb}
\usepackage{cite,url}

\def\q{\mbox{\boldmath $q$}}

\begin{document}

\title*{Relativistic Models of Quasielastic Electron and Neutrino-Nucleus
Scattering}

\author{\underline{{C.~Giusti}}\inst{1}
\and {A.~Meucci}\inst{1}
\and {F.~D.~Pacati}\inst{1}
\and {J.~A.~Caballero}\inst{2}
\and {J.~M.~Ud\'{\i}as}\inst{3}}

\institute{{Dipartimento di Fisica Nucleare e Teorica, 
Universit\`{a} degli Studi di Pavia and 
Istituto Nazionale di Fisica Nucleare, 
Sezione di Pavia, I-27100 Pavia, Italy}
\and {Departamento de F\'\i sica At\'omica, Molecular y
Nuclear, Universidad de Sevilla, E-41080 Sevilla, Spain}
\and {Grupo de F\'{\i}sica Nuclear, Departamento de F\'\i sica At\'omica, Molecular y
Nuclear, Universidad Complutense de Madrid, E-28040 Madrid, Spain}}

\maketitle

\begin{abstract}
Relativistic models developed for the exclusive and inclusive quasielastic (QE) 
electron scattering have been extended to charged-current (CC) and 
neutral-current (NC) neutrino-nucleus scattering. Different descriptions of
final-state interactions (FSI) are compared. For the
inclusive electron scattering the relativistic Green's function approach is compared with 
calculations based on the use of relativistic purely real mean field potentials in the final state.
Both approaches lead to a redistribution of the strength but conserving the total flux.
Results for the differential cross section at different energies 
are presented. Scaling properties are also analyzed and discussed.
\end{abstract}

\section{Introduction}

Electron scattering reactions with nuclei have provided the most detailed and
complete information on nuclear and nucleon structure \cite{book}. 
Additional information is available 
from  neutrino-nucleus scattering. Neutrinos can excite nuclear modes 
unaccessible in electron scattering, can give information on the hadronic weak 
current and on the strange  nucleon form factors. Although of great 
interest, such studies are not the only aim of many neutrino experiments, 
which are better aimed at a precise determination 
of neutrino properties. In neutrino oscillation experiments nuclei
are used to detect neutrinos. A proper analysis of data requires that the  
nuclear response to neutrino interactions is well under control and that the  
unavoidable theoretical uncertainties on nuclear effects are reduced as much as possible.

Different models developed and successfully tested in comparison
with electron scattering data have been extended to neutrino-nucleus scattering. 
Although the two situations are different, electron scattering is the best 
available guide to determine the prediction power of a nuclear model. 
Nonrelativistic and relativistic models have been used to describe nuclear
effects with different approximations. Relativity is important at all energies, 
in particular at high energies, and in the energy regime of many neutrino 
experiments a fully relativistic approach is required. 

Relativistic models for the exclusive and inclusive electron and neutrino
scattering in the QE region are presented in this contribution. In the QE 
region the nuclear response is dominated by one-nucleon knockout processes, where the probe interacts with a
quasifree nucleon that is emitted from the nucleus with a direct one-step
mechanism and the remaining nucleons are spectators.
In electron scattering experiments the outgoing nucleon can be detected in
coincidence with the scattered electron. In the exclusive $(e,e'p)$ reaction the
residual nucleus is left in a specific discrete eigenstate and the
final state is completely specified. In the inclusive $(e,e')$ scattering the 
outgoing nucleon is not detected and the cross section includes all the available
final nuclear states.

For an incident neutrino (antineutrino) NC and CC scattering can be considered  
\begin{eqnarray}
\nu (\bar\nu) + A & \rightarrow & {\nu'} (\bar\nu') + N + 
( A - 1)  \hspace{3cm} \mathrm{NC} \nonumber \\
\nu (\bar\nu) + A & \rightarrow & l^{-} (l^{+}) +
p(n) + (A-1) \hspace{2.47cm} \ \mathrm{CC} 
 \nonumber\end{eqnarray} 
In NC scattering only the emitted nucleon can be detected and the cross 
section is integrated over the energy and angle of the final lepton. Also 
the state of the residual $(A-1)$-nucleus is not determined and the cross 
section is summed over all the available final states. The 
same situation occurs for the CC reaction if only the outgoing nucleon is 
detected. The cross sections are therefore semi-inclusive in the hadronic 
sector and inclusive in the leptonic one and can be treated as an $(e,e'p)$ 
reaction where only the outgoing proton is detected. The exclusive CC process 
where the charged final lepton is detected in 
coincidence with the emitted nucleon can be considered as well. 
The inclusive CC scattering where only the charged lepton is detected 
can be treated with the same models used for the inclusive $(e,e')$ reaction. 

For all these processes the cross section is obtained  in the one-boson 
exchange approximation from the contraction between the lepton tensor, that
depends  only on the lepton kinematics,  and the hadron tensor $W^{\mu\nu}$,
that contains the nuclear response and whose components are given by bilinear 
products of the matrix elements of the nuclear current  $J^{\mu}$ between the 
initial and final nuclear states, i.e.,
\begin{equation}
W^{\mu\nu} = \sum_f \, \langle \Psi_f\mid J^{\mu}(\q) \mid \Psi_i\rangle \, 
\langle \Psi_i \mid J^{\nu\dagger}(\q)\mid \Psi_f\rangle \, 
\delta(E_i+\omega-E_f),
\label{eq.wmn}
\end{equation}
where $\omega$ and $\q$ are the energy and momentum transfer, respectively.
Different but consistent models to calculate $W^{\mu\nu}$ in QE electron and
neutrino-nucleus scattering are outlined in the next sections.

\section{Exclusive one-nucleon knockout}

Models based on the relativistic distorted-wave impulse approximation (RDWIA) 
have been developed \cite{meucci1,Ud} to describe the exclusive reaction 
where the outgoing nucleon is detected in coincidence with the 
scattered lepton and the residual nucleus is left in a discrete eigenstate 
$n$. In RDWIA the amplitudes of~(\ref{eq.wmn}) are obtained in a 
one-body representation as 
\begin{equation}
\langle\chi^{(-)}\mid  j^{\mu}(\q)\mid \varphi_n \rangle  \ ,
\label{eq.dko}
\end{equation}
where $\chi^{(-)}$ is the single- particle (s.p.) scattering state of the emitted 
nucleon, $\varphi_n$ is the overlap between the ground state of the target 
and the final state $n$, i.e., a s.p. bound state, and 
$j^{\mu}$  is the one-body nuclear current. In the model the s.p. bound and
scattering states are consistently derived as eigenfunctions of a Feshbach-type
optical potential \cite{book,meucci1}. Phenomenological ingredients are adopted
in the calculations. The bound states are Dirac-Hartree solutions 
of a Lagrangian, containing scalar and vector potentials, obtained in the 
framework of the relativistic mean-field theory  \cite{adfx,adfxa,adfxb}. The scattering
state is calculated solving the Dirac equation with relativistic 
energy-dependent complex optical potentials \cite{chc}.
RDWIA models have been quite successful in describing  a large amount of data 
(cross sections, response functions and polarization observables) for the 
exclusive  $(e,e^{\prime}p)$  reaction \cite{book,meucci1,Ud}. 

\section{Semi-inclusive neutrino-nucleus scattering}

The transition amplitudes of the NC and CC processes where only the outgoing 
nucleon is detected are described as the sum of the RDWIA amplitudes 
in~(\ref{eq.dko})  over the states $n$. 
In the calculations \cite{nc} a pure shell-model (SM) description is assumed, i.e., 
$n$  is a one-hole state and the sum is over all the occupied SM states. FSI 
are described by a complex optical potential whose imaginary part gives an
absorption that reduces the calculated cross section. A similar reduction is 
found in $(e,e^{\prime}p)$ calculations. 
The imaginary part accounts for the flux lost in a specific channel towards 
other channels.
This approach is conceptually correct for an exclusive reaction, where only 
one channel contributes, but it would be wrong for the inclusive scattering, 
where all the channels contribute and the total flux must be conserved. For the 
semi-inclusive process where an emitted nucleon is detected, some of the 
reaction channels which are responsible for the imaginary part of the potential, 
are not included in the experimental cross section and it is correct to include 
in the calculations the absorptive imaginary part. There are, however, 
contributions that are not included in this model and that can be included in 
the experimental cross section, for instance, contributions due to multi-step 
processes. 
The relevance of these contributions depends on kinematics 
and should not be large in the QE region.

\section{Inclusive lepton-nucleus scattering}

In the inclusive scattering where only the outgoing lepton is
detected FSI are treated in the Green's function (GF) approach  
\cite{ee,eea,cc,eenr}. In this model the components of the hadron tensor are 
written in terms of the s.p. optical model Green's function. This is the result 
of suitable approximations, such as the assumption of a one-body current and 
subtler approximations related to the IA. 
The explicit calculation of the s.p. Green's function is avoided by its 
spectral representation, which is based on a biorthogonal expansion in terms of 
a non Hermitian optical potential $\cal H$ and of its Hermitian conjugate 
$\cal H^{\dagger}$. Calculations require matrix elements of 
the same type as the RDWIA ones in Eq. \ref{eq.dko}, but involve 
eigenfunctions of both $\cal H$ and $\cal H^{\dagger}$, where the imaginary 
part gives in one case an absorption and in the other case a gain of flux, and 
in the sum over $n$ the total flux is redistributed and conserved.  
The GF approach guarantees a consistent treatment of FSI in the exclusive and in 
the inclusive scattering and gives a good description of $(e,e')$ data \cite{ee}.

\begin{figure}[t]\centering
\includegraphics[width=110mm,height=70mm]{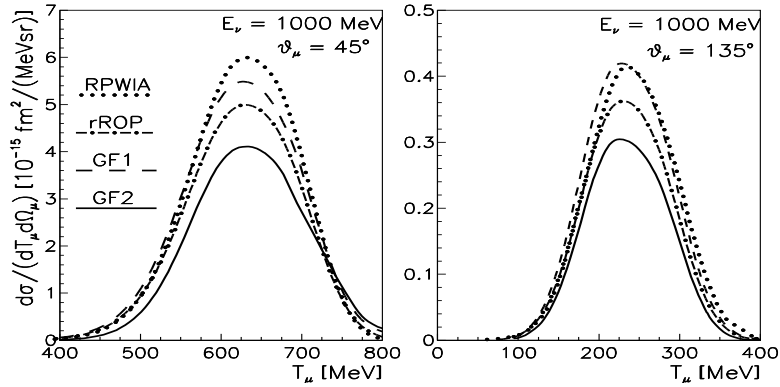}
\vskip  -13mm
\caption{The cross sections of the $^{12}$C$(\nu_{\mu},\mu^-)$ 
reaction for an incident neutrino energy of $E_\nu$ = 1000 MeV and a muon
scattering angle $\theta_\mu = 45^{\mathrm o}$ (left panel) 
and $45^{\mathrm o}$ (right panel) as a
function of the muon kinetic energy $T_\mu$.  
Results for RPWIA (dotted), rROP (dot-dashed), and GF with  two 
different optical potentials \cite{chc}, i.e., EDAD1 (GF1) and EDAD2 (GF2), are 
compared. 
\label{fig1}}
\end{figure}

In Fig. \ref{fig1} the $^{12}$C$(\nu_{\mu},\mu^-)$  cross sections calculated
with the GF approach and two parametrizations of the 
optical potential \cite{chc}, i.e., EDAD1 (GF1) and EDA2 (GF2), are compared
with the results of the relativistic plane wave IA (RPWIA), where FSI 
are neglected. The cross sections obtained when only the real part of the 
relativistic optical potential (rROP) is retained and the imaginary part is 
neglected are also shown in the figure. This approximation conserves the flux, 
but it is conceptually wrong because the optical potential has to be complex 
owing to the presence of inelastic channels. The differences between the rROP, 
GF1, and GF2 results are due to the imaginary part of the optical potential. 
Different parameterizations give similar real terms and the rROP cross sections 
are practically insensitive to the choice of the optical potential. In contrast, 
the imaginary part is sensitive to the parameterization of the ROP and gives 
the differences between the rROP and GF results, but also between the two GF
calculations shown in the figure.

The analysis of data for neutrino experiments requires a precise knowledge of 
lepton-nucleus cross sections, where uncertainties on nuclear effects are 
reduced as much as possible. To this aim, it is important to check the 
consistency of different models and the validity of the adopted approximations.

\begin{figure}[t]\centering
\includegraphics[width=100mm]{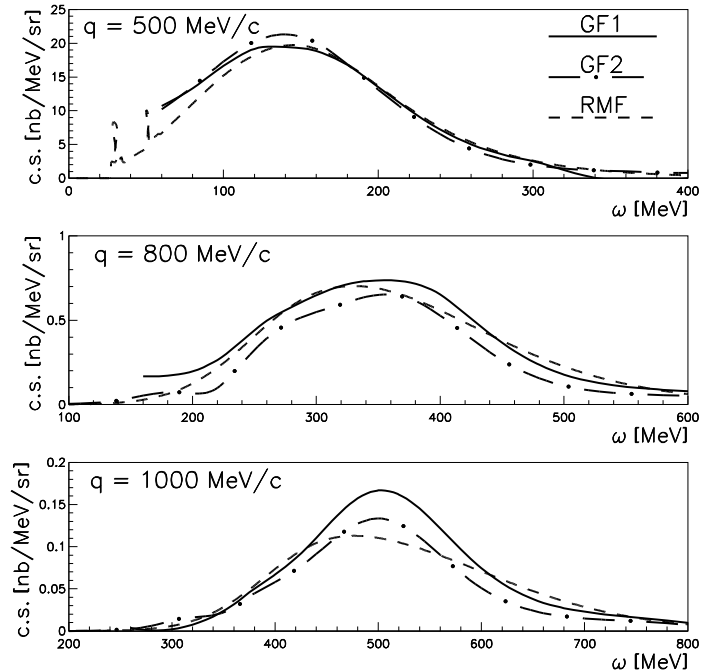}
  \caption{The differential cross sections of the $^{12}$C$(e,e')$ 
reaction for an incident electron energy  of 1 GeV and three values of the 
momentum transfer, i.e., $q = $ 500, 800, and 1000 MeV$/c$. 
Line convention: GF1 (solid), GF2 (long dot-dashed), RMF (dashed). 
\label{fig2}}
\end{figure}
\begin{figure}[t]\centering
\includegraphics[width=100mm]{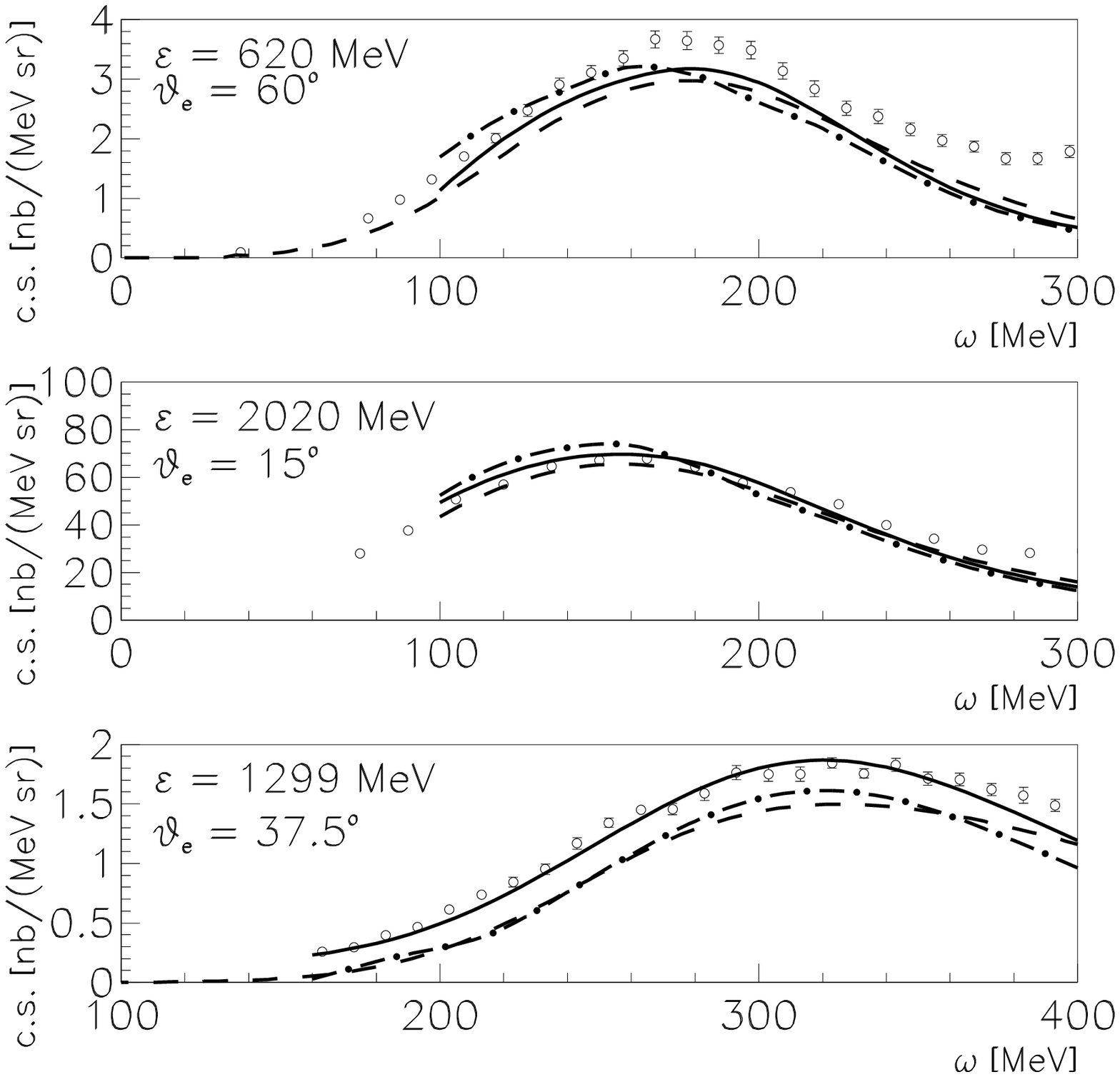}
  \caption{Differential cross section of the $^{12}$C$(e,e^{\prime})$ reaction 
for different beam energies and electron scattering angles. Line convention as in 
Fig. \ref{fig2}, experimental data from \cite{620,day,1299}.
\label{fig3}}
\end{figure}

The results of the relativistic models developed by the Pavia and the 
Madrid-Sevilla groups for the inclusive electron scattering are
compared in \cite{comp}.
As a first step the consistency of the RPWIA and rROP calculations performed 
by the two groups with independent  numerical programs has been 
checked. Then the results of different descriptions of FSI have been compared. 
An example is shown in  Fig. \ref{fig2}, where the  $^{12}$C$(e,e')$ cross 
sections obtained with the GF approach and the two parametrizations of the 
optical potential already used in Fig. \ref{fig1} are compared
with the results of the relativistic mean  field (RMF) model \cite{cab}, where 
the scattering wave functions are calculated with the same real potential used 
for the initial bound states. The RMF model fulfills the dispersion relation and maintains 
the continuity equation.
The differences between the RMF and GF results increase with the momentum 
transfer. Also the discrepancies between the GF1 and GF2 cross sections depend 
on the momentum transfer.  At $q$ = 500 MeV$/c$ the three results are similar, 
both in magnitude and shape. Moderate differences are found at $q$ = 800 
MeV$/c$, and larger differences at $q = 1000$ MeV$/c$.
The shape of the RMF cross section shows an asymmetry, with a long tail 
extending towards higher values of $\omega$, that is essentially due to the 
strong energy-independent scalar and vector potentials present in the RMF approach.
The asymmetry towards higher $\omega$ is less significant but still visible 
for GF1 and GF2, whose cross sections show a
similar shape but with a significant difference in the magnitude. 
At $q$ = 1000 MeV$/c$ both GF1 and GF2 cross sections are higher than the RMF 
one in the region where the maximum occurs.
A stronger enhancement is obtained with GF1, which at the peak 
overshoots the RMF cross section up to $40\%$.

The behaviour of the RMF and GF results as a function of $q$ and
$\omega$  is linked to the structure of the relativistic 
potentials involved in the RMF and GF models. Whereas RMF is based on the 
use of a strong energy-independent real potential, GF makes use 
of a complex energy-dependent optical potential. In GF calculations the 
behavior of the optical potential changes with the momentum and energy 
transferred in the process, and higher values 
of $q$ and $\omega$ correspond to higher energies for the optical potential. 
The GF results are consistent with the general behavior of the 
optical potentials and are basically due to their imaginary part, that includes the overall effect of the
inelastic channels and is not univocally determined by the elastic phenomenology.  
Different parameterizations give similar real terms and the rROP cross sections 
are practically insensitive to the choice of the optical potential. 
The real part decreases increasing the 
energy and the rROP result approaches the RPWIA one for large values of 
$\omega$ \cite{comp}. 
In contrast, the imaginary part has its maximum strength around 500 MeV and is
sensitive to the parameterization of the ROP. 

In Fig. \ref{fig3} the GF1, GF2, and RMF results are compared with the
experimental cross sections for three different kinematics. 
The three models lead to similar cross sections. The main differences are presented 
for higher values of $q$, about 800 MeV/$c$ (bottom panel),
where the GF1 cross section is larger than the GF2  
and RMF ones. The experimental cross section is slightly underpredicted 
in the top panel and well described in the middle panel by all calculations. 
The results in the bottom panel show a fair agreement with data for
GF1, whereas GF2 and RMF underpredict the experiment.
Although satisfactory on general grounds, the comparison with data gives here only an indication and cannot be conclusive 
until contributions beyond the QE peak, like meson exchange currents 
and $\Delta$ effects, which may play a significant role in the analysis of data 
even at the maximum of the QE peak, are carefully 
evaluated \cite{BCDM04,amaro05,ivanov08}.

\section{Scaling functions}

The comparison between the results of the Pavia and Madrid-Sevilla groups has 
been extended to the scaling properties of the different
relativistic models \cite{comp}.

Scaling ideas applied to inclusive QE electron-nucleus scattering
have been shown to work properly to high accuracy \cite{mai1,don1}.
At sufficiently high momentum transfer a scaling function is derived dividing 
the experimental $(e,e^{\prime})$ cross sections by an appropriate 
single-nucleon cross section. This is basically the idea of the IA. If this 
scaling function depends only upon one kinematical variable, the scaling 
variable, one has scaling of first kind. If the scaling function is roughly the 
same for all nuclei, one has scaling of second kind. When both kinds of scaling 
are fulfilled, one says that superscaling occurs.   
An extensive analysis of electron scattering data has shown 
that scaling of first kind is fulfilled at the left of the QE peak and broken 
at its right, whereas scaling of second kind is well satisfied at the
left of the peak and not so badly violated at its right. A phenomenological 
scaling function $f_L^{exp}(\psi')$ has been extracted from data of the 
longitudinal response in the QE region. The dimensioneless scaling
variable $\psi'(q,\omega)$ is extracted from the relativistic Fermi gas (RFG) 
analysis that incorporates the typical momentum scale for the selected
nucleus \cite{mai1,cab1}.
Although many models based on the IA
exhibit superscaling, even perfectly as the RFG, only a few
of them are able to reproduce the asymmetric shape of $f_L^{exp}(\psi')$ with a significant
tail extended to high values of $\omega$ (large positive values of $\psi'$).
One of these is the RMF model where FSI are described by the same real relativistic 
potential used for the initial bound states. In contrast, the RPWIA and rROP, 
although satisfying superscaling
properties, lead to symmetrical-shape scaling functions which are not in 
accordance with data analysis \cite{cab1,cab2}.

\begin{figure}[t]\centering
\includegraphics[width=100mm]{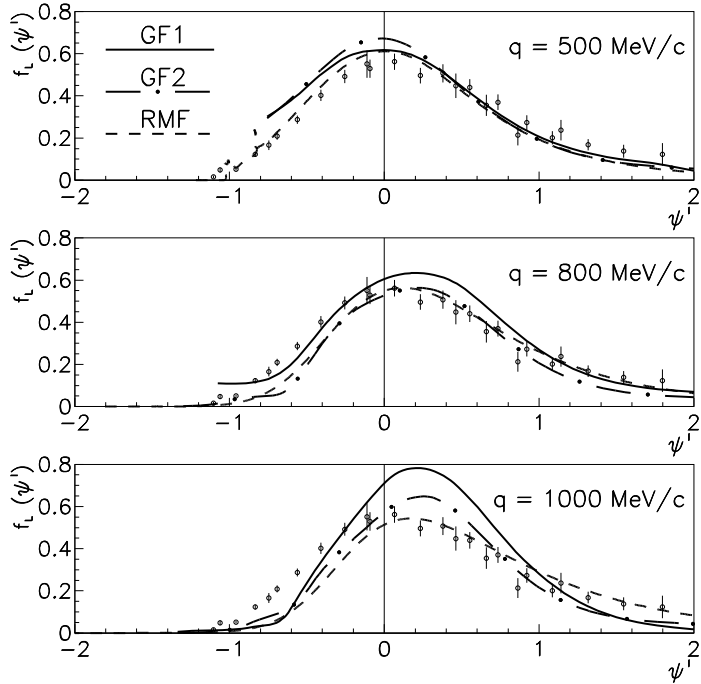}
  \caption{Longitudinal contribution to the scaling function for  $q = $ 500, 
800, and 1000 MeV$/c$  with the GF1 (solid), GF2 
(long dot-dashed), and RMF (dashed) models compared with the averaged 
experimental scaling function.
\label{fig4}}
\end{figure}
In Fig. \ref{fig4} the scaling function $f_L(\psi')$ evaluated with RMF, 
GF1, and GF2 for different values of $q$ are compared to the
phenomenological function  $f_L^{exp}(\psi')$. The RMF model produces an 
asymmetric shape with a long tail in the region with $\psi^{\prime}>0$ that 
follows closely the phenomenological function behavior. The GF results are
similar to the RMF ones at $q$ = 500 MeV$/c$ and, with moderate differences, at
$q$ = 800 MeV$/c$, while visible discrepancies appear at $q$ = 1000 MeV$/c$.
The discussion is similar to the one applied to the 
cross sections in Fig. \ref{fig3}. The asymmetric shape with a 
tail in the region of  positive $\psi^{\prime}$ is obtained with both RMF and GF. 
The different dependence on $q$ shown by the potentials 
involved in RMF and GF makes the tail of the GF scaling function less 
pronounced as the value of $q$ goes up.

Except for the highest value of $q$ considered 
(1000 MeV/$c$), GF1, GF2. and RMF yield very similar predictions for
$f_L(\psi')$, in good agreement with the experimental function. 
The asymmetric tail of the data and the strength at the peak are 
fairly reproduced by the three approaches. For $q=1000$ MeV/$c$, however, only
RMF seems to be favoured from the comparison to data, while GF1 and GF2 
yield rather different predictions than RMF, that seem to be ruled out by
data. We note that as the momentum transfer increases the phenomenological 
optical potentials, that is used as input in the GF approach, will (implicitely) 
incorporate a larger amount of contributions from non 
nucleonic degrees of freedom, such as, for instance, the loss of (elastic) flux
into the inelastic $\Delta$ excitation with or without real pion production. Thus 
the input of the GF formalism is contaminated 
by non purely nucleonic contributions. As a consequence, GF predictions
depart from the experimental QE longitudinal response, that
effectively isolates only nucleonic contributions. This difference,
which increases with increasing $q$, emerges
as an excess of strength predicted by the GF model as it translates a loss of 
flux due to non-nucleonic processes into inclusive purely nucleonic strength.
On the other hand, the RMF model uses as input the effective mean field that reproduces
the saturation properties of nuclear matter and of the ground state of the nuclei involved, and it is more suited to
estimate the purely nucleonic contribution to the inclusive cross-section, 
even at $q=1000$ MeV/$c$.

\section{Summary and conclusions}

Relativistic models developed for QE electron scattering and successfully tested
in comparison with experimental data have been extended to calculate CC and NC
neutrino-nucleus cross sections. In the models nuclear effects are treated
consistently  in exclusive, semi-inclusive, and inclusive reactions. 

The comparison of different models is important to reduce theoretical 
uncertainties on nuclear effects. 
The results of the relativistic models developed by the Pavia and the 
Madrid-Sevilla groups to describe FSI in the inclusive QE electron-nucleus 
scattering have been compared. The consistency of the calculations of the two 
groups has been checked in RPWIA and rROP. Then two different models based on 
the RIA have been compared: the GF approach of the Pavia group, that is based 
on the use of the complex relativistic optical potential and which allows one 
to treat FSI consistently in the inclusive and exclusive reactions, and the RMF
model of the Madrid-Sevilla group, where the distorted waves are obtained with 
the same real relativistic mean field considered for the bound states. 
Results are compared for the differential cross sections and scaling functions.
Discrepancies increase with the momentum transfer. This is linked to the 
energy-dependent optical potentials involved in the GF method by contrast to 
the energy-independent RMF potentials. 

All models considered respect scaling and superscaling properties.
The significant asymmetry in the scaling function produced by the RMF model 
is strongly supported by data \cite{cab1}. The relativistic GF approach leads 
to similar results to RMF, i.e., with the asymmetry for intermediate $q$-values.
Visible discrepancies emerge for larger $q$, being the GF scaling function tail 
less pronounced but showing more strength in the maximum region. Moreover, the 
GF results for high $q$ present a strong dependence on the parameterization of 
the optical potential, in particular of its imaginary part.  
The GF approach, even based on the use of a complex optical 
potential, conserves the flux and the imaginary term redistributes the strength 
among different channels. This explains the difference observed between RMF and 
GF predictions, the latter with additional strength in the region close to the 
maximum in the QE response.This behavior could be connected with effects coming 
from the contribution of the $\Delta$ which are, somehow, accounted for in a 
phenomenological way by the GF approach, modifying the responses even in the 
region where the QE peak gives the main contribution. 
We must keep in mind that the higher the momentum transfer, the stronger 
the overlap between the QE and $\Delta$ peaks, and it is very difficult to 
isolate contributions coming from either region.

The similarities of the GF and RMF predictions, particularly for intermediate 
values of $q$, and the very reasonable agreement with the data for the 
longitudinal scaled response are a clear indication that both models make a very decent job 
in estimating the inclusive contribution.

\end{document}